\documentclass[twocolumn,english,aps,prb,showpacs,byrevtex,amsmath,amssymb,superscriptaddress,longbibliography]{revtex4-1}
\usepackage[T1]{fontenc}
\usepackage[latin9]{inputenc}
\usepackage{lineno}
\usepackage{textcomp}
\usepackage{amsmath}
\usepackage{graphicx}
\usepackage{multirow}
\usepackage{ bbold }
\usepackage{lipsum}
\usepackage{array}
\newcolumntype{P}[1]{>{\centering\arraybackslash}p{#1}}

\makeatletter
\usepackage{lipsum}
\usepackage{epsfig}

\usepackage[colorlinks,linkcolor=blue]{hyperref}
\usepackage[nameinlink,capitalise]{cleveref}
\usepackage{subfigure}
\usepackage{color}
\usepackage{physics}
\usepackage[title]{appendix}

\definecolor{Red}{rgb}{1,0,0}
\definecolor{Blu}{rgb}{0,0,1}
\definecolor{Green}{rgb}{0,1,0}

\makeatother
\usepackage{amssymb}
\usepackage{babel}

\usepackage{tikz,xcolor,hyperref}
\definecolor{lime}{HTML}{A6CE39}
\DeclareRobustCommand{\orcidicon}{%
	\begin{tikzpicture}
	\draw[lime, fill=lime] (0,0)
	circle [radius=0.16]
	node[white] {{\fontfamily{qag}\selectfont \tiny ID}};
	\draw[white, fill=white] (-0.0625,0.095)
	circle [radius=0.007];
	\end{tikzpicture}
	\hspace{-2mm}
}

\foreach \x in {A, ..., Z}{%
	\expandafter\xdef\csname orcid\x\endcsname{\noexpand\href{https://orcid.org/\csname orcidauthor\x\endcsname}{\noexpand\orcidicon}}
}

\begin{document}

\title{Engineering axion insulator and other topological phases\\ in superlattices without inversion symmetry}


\author{Rajibul Islam\orcidC}
\email{rislam@magtop.ifpan.edu.pl}
\affiliation{International Research Centre MagTop, Institute of Physics, Polish Academy of Sciences, Aleja Lotnik\'ow 32/46, PL-02668 Warsaw, Poland}

\author{Sougata Mardanya}
\affiliation{Department of Physics, National Cheng Kung University, Tainan 70101, Taiwan}

\author{Alexander Lau\orcidE}
\affiliation{International Research Centre MagTop, Institute of Physics, Polish Academy of Sciences, Aleja Lotnik\'ow 32/46, PL-02668 Warsaw, Poland}

\author{Giuseppe Cuono\orcidD}
\affiliation{International Research Centre MagTop, Institute of Physics, Polish Academy of Sciences, Aleja Lotnik\'ow 32/46, PL-02668 Warsaw, Poland}

\author{Tay-Rong Chang\orcidH}
\affiliation{Department of Physics, National Cheng Kung University, Tainan 70101, Taiwan}

\author{Bahadur Singh\orcidG}
\email{bahadur.singh@tifr.res.in}
\affiliation{Department of Condensed Matter Physics and Materials Science,
Tata Institute of Fundamental Research, Colaba, Mumbai 400005, India}

\author{Carlo M. Canali\orcidF}
\affiliation{Department of Physics and Electrical Engineering, Linnaeus University, 392 31 Kalmar, Sweden}

\author{Tomasz Dietl\orcidB}
\affiliation{International Research Centre MagTop, Institute of Physics, Polish Academy of Sciences, Aleja Lotnik\'ow 32/46, PL-02668 Warsaw, Poland}
\affiliation{WPI-Advanced Institute for Materials Research, Tohoku University, Sendai 980-8577, Japan}

\author{Carmine Autieri\orcidA}
\email{autieri@magtop.ifpan.edu.pl}
\affiliation{International Research Centre MagTop, Institute of Physics, Polish Academy of Sciences, Aleja Lotnik\'ow 32/46, PL-02668 Warsaw, Poland}
\affiliation{Consiglio Nazionale delle Ricerche CNR-SPIN, UOS Salerno, I-84084 Fisciano (Salerno),
Italy}

\begin{abstract}
We study theoretically the interplay between magnetism and topology in three-dimensional HgTe/MnTe superlattices stacked along the (001) axis. Our results show the evolution of the magnetic topological phases with respect to the magnetic configurations. An axion insulator phase is observed for the antiferromagnetic order with the out-of-plane N\'eel vector direction below a critical thickness of MnTe, which is the ground state amongst all magnetic configurations.
Defining $T$ as the time-reversal symmetry, this axion insulator phase is protected by a magnetic two-fold rotational symmetry $C_2{\cdot}T$.
We find that the axion insulator phase evolves into a trivial insulator as we increase the number of the magnetic MnTe layers, and we present an estimate of the critical thickness of the MnTe film above which the axion insulator phase is absent.
By switching the N\'eel vector direction into the $ab$ plane, the system realizes different antiferromagnetic topological insulators depending on the thickness of MnTe.
These phases feature gapless surface Dirac cones shifted away from high-symmetry points on surfaces perpendicular to the N\'eel vector direction of the magnetic layers.
In the presence of ferromagnetism, the system realizes a magnetic Weyl semimetal and a ferromagnetic semimetal for out-of-plane and in-plane magnetization directions, respectively.
We observe large anomalous Hall conductivity in the presence of ferromagnetism in the three-dimensional superlattice.

\end{abstract}
\date{\today}
\maketitle

\section{Introduction}
In modern condensed matter physics, topological phases have gained attention for their exotic properties \cite{bansil2016colloquium,Brzezicki_2019}.
Symmetries play an important role in the protection of the topological phases. One of the most important symmetries is the time-reversal symmetry (TRS). Starting from a system with TRS, this symmetry can be broken by doping with magnetic impurities or by creating heterostructures and superlattices between magnetic and topological materials.
Breaking of the TRS can lead to exotic phases such as the quantum anomalous Hall (QAH) phase  \cite{AQHE_theory_Mn,Yu:2010_S,AQHE_experimentalBi2Se3,Chang2022Colloquium}, the antiferromagnetic topological insulator (AFM TI) \cite{Rienks2019}, magnetic Weyl\cite{wadge2021electronic} and Dirac semimetals, or nodal-line semimetal phases.\cite{Bernevig2022}
Besides that, also classical properties, such as the anomalous Hall effect, are affected by topology \cite{vanThiel21coupling,Groenendijk20Berry,Brzezicki22}. On the other hand,  the Van Vleck-Bloembergen-Rowland mechanism, thought to lead to ferromagnetism in magnetic topological materials  \cite{Yu:2010_S},  appears to be dominated by superexchange \cite{sliwa2021superexchange}.

In the case of broken TRS, one of the most studied phase is the axion insulator\cite{Sekine2021} (AXI), especially in the context of Bi$_2$Te$_3$-based compounds doped with Cr \cite{Mogi2022}, with interstitial Mn  \cite{Zhang2019,Liu2020,Gao2021}, or interfaced with other antiferromagnetic transition metal compounds \cite{Pournaghavi2021Realization}. Significant experimental effort has been made in this direction to realize the AXI phase, which had previously been predicted theoretically.
The AXI phase has gapped surfaces exhibiting a topologically protected half-quantized surface anomalous Hall conductivity perpendicular to the surface. The sign of the half-quantized surface anomalous Hall conductivity depends on the specific surface termination.
In most cases, the AXI is realized by an AFM phase with half-quantized anomalous Hall conductivity on opposite surfaces, whose contributions cancel each other out.
A clear half-quantized anomalous Hall conductivity has also been found in the ferromagnetic axion insulators MnBi$_8$Te$_{13}$  \cite{doi:10.1126/sciadv.aba4275} and in a heterostructure of ferromagnetic (FM) Cr-doped (Bi,Sb)$_2$Te$_3$ and non-magnetic TI (Bi,Sb)$_2$Te$_3$  \cite{Mogi2022}.
The debate on the dimensionality of the axion insulator is still open  \cite{Liu2020}.
Importantly, the axion insulator phase must involve a material that is a three-dimensional topological insulator \cite{Fijalkowski2021any,PhysRevLett.118.246801}.

Generally, magnetic topological insulators are three-dimensional magnetic materials with a non-trivial topological index protected by a symmetry other than TRS.
In the past years, there have been many studies in which the nontrivial $\mathbb{Z}_2$ index of the axion insulator is protected by inversion symmetry \cite{10.21468/SciPostPhys.6.4.046,Varnava2018surfaces}. On the materials side, the axion phase protected by inversion symmetry has been widely investigated in the MnTe(Bi$_2$Te$_3$)$_n$ material class \cite{doi:10.1126/sciadv.aaw5685,doi:10.1063/5.0059447,PhysRevLett.122.206401,Liu2020,Klimovskikh2020}. However, a few cases have been proposed where the AXI phase is protected by other crystal symmetries \cite{Varnava2020axion,Riberolles2021magnetic,Bahadur22}.

In this paper, we construct a 3D superlattice (SL) without inversion symmetry composed of the HgTe topological zero-gap insulator and the cubic MnTe antiferromagnetic insulator.
We define $C_{2}$ and $T$ as the two-fold rotational symmetry and the time-reversal symmetry, respectively.
We show that the three-dimensional HgTe/MnTe superlattice realizes an AXI ground state protected by a magnetic two-fold rotational symmetry $C_{2}{\cdot}T$ symmetry for a MnTe-layer thickness of two unit cells (uc).
By increasing the MnTe thickness to four unit cells, the 3D HgTe/MnTe SL becomes a trivial insulator. While our focus is on the ground state of the superlattice, we also discuss other topological phases arising for different magnetic configurations.
All calculations are performed within density functional theory (DFT), whose implementation is presented in Appendix A.

HgTe is a symmetry-enforced zero-gap semiconductor with zinc-blende crystal structure and a band inversion at the $\Gamma$ point.
Both in the bulk and in superlattices, the \emph{s}-band composed of the bonding state of \emph{s}-orbitals of Hg goes below the top of the bands composed of the \emph{p}-orbitals of Te \cite{ortix2014absence,kirtschig2016surface,autieri2020momentumresolved}.
Therefore, we have a single band inversion making the topology of the HgTe-based supercell robust \cite{Islam22}, while avoiding the chance of double band inversions.
Many studies on HgTe doped with magnetic impurities suggest the presence of a quantum anomalous Hall phase \cite{liu2008quantum,Yu2010quantized,chang2013experimental,HgTe_QAHE,Cuono22} in a suitable range of magnetic doping where the band inversion is still present.

The Weyl semimetal (WSM) phase was proposed \cite{HgTe_weyl} and studied experimentally \cite{Mahler:2019_PRX} in HgTe epilayers under biaxial compressive strain and in the presence of the Dresselhaus non-local spin-orbit coupling.
Other theoretical investigations proposed the Weyl phase in bulk zinc-blende compounds in the presence of first-neighbor spin-orbit coupling between \emph{s}-Hg and \emph{p}-Te \cite{PhysRevB.87.245112}.
There are no further investigations on the correct symmetry of the first-neighbor spin-orbit coupling in zinc-blende crystals. Using the wannierization method, we have recently found that a strong second-neighbour spin-orbit coupling between extended \emph{p}-Te orbitals is present in the bulk.
However, the symmetry makes the spin-orbit couplings ineffective for generating a Weyl phase \cite{HgTe_weyl}.
Nevertheless, through breaking of the crystal symmetry by strain or by creating a superlattice, this second-neighbour spin-orbit coupling can lead to a Weyl phase \cite{Islam22}.
Therefore, also the MnTe/HgTe superlattice can stabilize the WSM phase thanks to the breaking of the crystal symmetry and to the non-local spin-orbit coupling term in the Hamiltonian.
Additionally, we have checked that this non-local spin-orbit coupling is also responsible for the triple point accompanied by a nodal-line phase when HgTe is strained along the (111) direction \cite{PhysRevX.6.031003}.
Different topological phases appear in the 2D limit as the quantum spin Hall phase in quantum wells \cite{doi:10.1126/science.1133734,Nguyen22} and
another Weyl phase \cite{Weyl2D}. These 2D phases will not be considered in this paper.\\

The remainder of the text is structured as follows. In Sec.~II, we report the crystal structure and symmetry of the considered systems.
The topological phases of the 3D SLs in the presence of antiferromagnetism are described in Sec.~III.
The ferromagnetic topological phases of 3D SLs are described in Sec.~IV.
Finally, we draw our conclusion in Sec.~V.

\section{Crystal structure and symmetry}

HgTe is an II-VI semiconductor with cubic zinc-blende crystal structure, which is the stable structural phase at ambient conditions. It belongs to the space group (SG) $F\bar{4}3m$, No.~216, with a mirror plane along the (011) direction.
It has eight atoms in the conventional unit cell with cations at the (0,0,0) position and anions at the $\frac{1}{4}$(1,1,1) position.
The structural ground state of bulk MnTe is the $\alpha$-phase with a hexagonal crystal structure \cite{PhysRevB.96.214418,Bossini_2020} and has recently attracted interest due to the presence of altermagnetism \cite{altermagnetism} below N\'eel temperature $T_{\text{N}}= 310$\,K  \cite{PhysRevB.96.214418}.
When MnTe is grown on cubic zinc-blende substrates, it can also assume the cubic zinc-blende phase \cite{SZUSZKIEWICZ1999425,Zakrzewski1995,PhysRevB.46.12076,PhysRevB.48.12817,doi:10.1063/1.116620}, which is the only MnTe phase investigated in this paper. Mn$^{2+}$ ions have a 3d$^5$ high-spin configuration  which is magnetically frustrated in the perfectly cubic zinc-blende structure by antiferromagnetic superexchange interactions that result in $T_\text{N}= 68$\,K  \cite{SZUSZKIEWICZ1999425}.
Bulk HgTe can realize a topological insulator under biaxial strain, whereas bulk MnTe is an AFM insulator. The lattice parameters of bulk HgTe and MnTe are $a_\mathrm{HgTe}= 6.46\,${\AA} and $a_\mathrm{MnTe}= 6.32\,${\AA}, respectively.

\begin{figure}
  \begin{center}
        \includegraphics[width=0.99\linewidth]{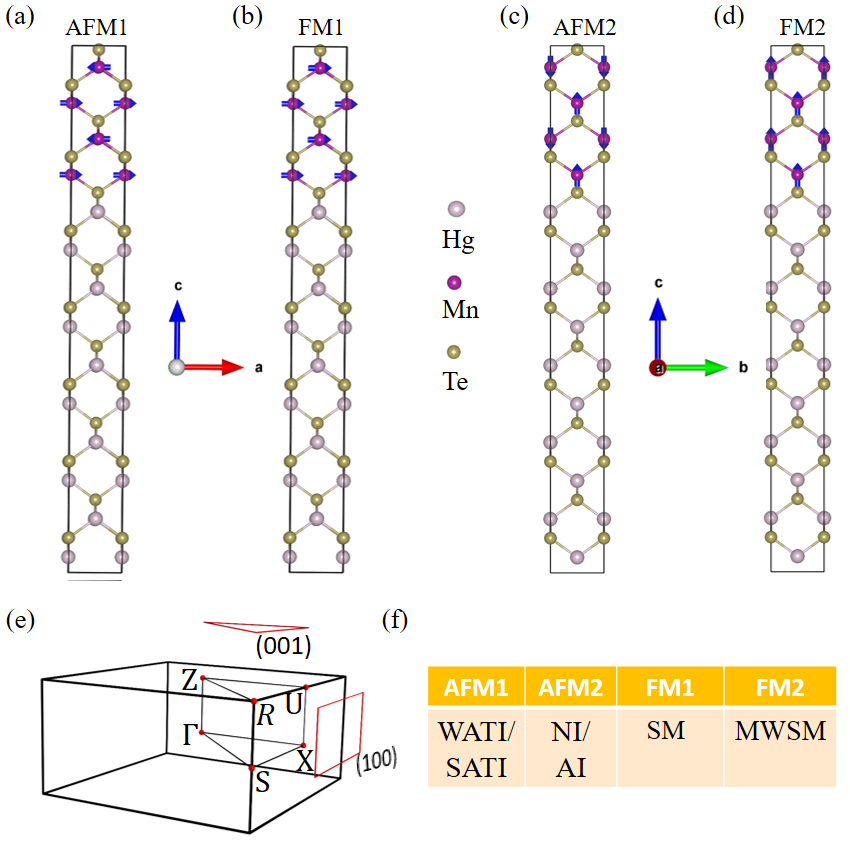}
  \end{center}
   \centering
   \caption{Crystal structure and magnetic configurations of 3D (HgTe)$_{10}$/(MnTe)$_4$ superlattices. Hg, Mn and Te atoms are represented by grey, purple and yellow balls, respectively. The bold black lines represent the unit cell of the superlattices. The spins of the Mn atoms are indicated as blue arrows. Panels (a-d) show the magnetic configurations AFM1, FM1, AFM2 and FM2, respectively. The in-plane magnetic moment is oriented along the $x$-axis. (e) Brillouin zone of the tetragonal 3D superlattice. The path used for the bulk band structure is in black, while the (100) and (001) surfaces are in red. (f) Topological phases associated with the different magnetic configurations. The acronyms are described in the text.}
   \label{bulk}
\end{figure}

Recent theoretical predictions show that HgTe-based 3D superlattices can host Weyl or nodal-line semimetal phases \cite{Islam22}. Going beyond this study, here we exploit a 3D superlattice of HgTe/MnTe with different magnetic configurations. HgTe and MnTe are vertically stacked along the (001) axis.
We analyze 3D superlattices of HgTe/MnTe that are magnetically symmetric with an even number of HgTe and MnTe layers. For the antiferromagnetic configurations, we will focus on two cases (HgTe)$_{10}$/(MnTe)$_4$ and (HgTe)$_{10}$/(MnTe)$_2$ that we will define as 4 MnTe uc and 2 MnTe uc, respectively. For the ferromagnetic configuration, we will just investigate the (HgTe)$_{10}$/(MnTe)$_4$ SL. Figures \ref{bulk}(a-d) show the crystal structure for the (HgTe)$_{10}$/(MnTe)$_4$ SL. In the figures where it is not otherwise specified, we refer to the case with 4 MnTe uc.
In the magnetically asymmetric case, the surfaces of these systems tend to become metallic and uncompensated Berry phases lead to a net anomalous Hall effect in the metallic cases \cite{asa2020anomalous}.
However, in the insulating case of a TI, asymmetric systems can generate the AXI phase \cite{pournaghavi2021TME} as well as the exotic phase of the ferromagnetic axion insulator \cite{doi:10.1126/sciadv.aba4275}.
Importantly for our discussion, the system does not have inversion symmetry.

\begin{figure}
  \begin{center}
\includegraphics[width=0.75\linewidth]{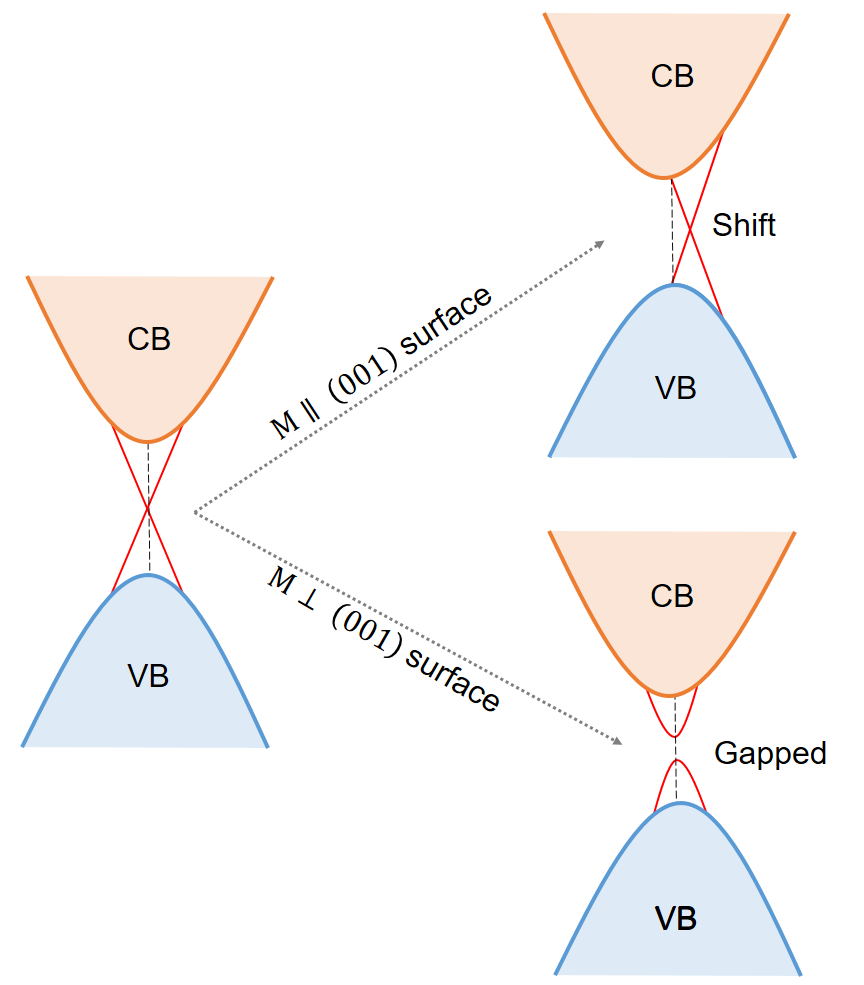}
  \end{center}
\centering
\caption{
Schematic representation depicting how the surface Dirac cone can change when magnetism is introduced in a topological insulator. The red lines represent topological surface states that, without magnetism, connect valence bands (VB) and conduction bands (CB) with a surface Dirac point pinned to a high-symmetry point. With magnetism, the Dirac cone is either shifted or gapped out paving the way to an antiferromagnetic topological insulator or to an axion insulator, respectively.
This representation does not apply to the ferromagnetic phases since they are semimetallic.
}
   \label{Schematic}
\end{figure}

Generally, for a topological insulator, a net magnetic moment on the surface can either gap out the surface Dirac cone or shift it away from high-symmetry points \cite{Menshov2021}, as illustrated schematically in \cref{Schematic}.
The exact mechanism is a result of the interplay between the considered surface termination, the magnetization or N\'eel vector direction, and the remaining symmetries of the material.
In particular, additional symmetries, such as magnetic rotation symmetries, may pin a surface Dirac cone to a high-symmetry point or line without gapping it out.
In the next subsections, we will show this interplay for the (100), (010), and (001) surface terminations of the topological insulator with different magnetization or N\'eel vector directions.

The strong topological insulator presents surface Dirac cones on all surfaces, while the weak topological insulator presents surface Dirac cones just on some surfaces.
For antiferromagnetic order with in-plane N\'eel vector, which is denoted as AFM1 in \cref{bulk}(a),
we find that the system is in a weak or strong antiferromagnetic topological insulating phase (WATI or SATI).
In the case of antiferromagnetic order with out-of-plane N\'eel vector, which is denoted as AFM2 in \cref{bulk}c, the system realizes either an AXI phase or a normal insulator (NI) phase.
In both cases, which particular phase is realized depends on the thickness of the MnTe layers. The increase in the thickness of the topologically trivial MnTe weakens the topological phase as we will show below.
Out-of-plane ferromagnetic order (FM2 in \cref{bulk}d) gives rise to a magnetic Weyl semimetal (MWSM) phase, whereas in-plane ferromagnetic order (FM1 in \cref{bulk}b) leads to a ferromagnetic semimetal (SM).
If the magnetic moments are in-plane, the (100) and (010) surfaces are inequivalent. On the contrary, when the magnetic moments are out-of-plane, the (100) and (010) surfaces are equivalent.

The in-plane lattice constant of the supercell depends on the lattice constants of the constituent compounds.
Since the lattice constant of HgTe is larger than MnTe, the superlattice produces an in-plane tensile strain on MnTe. The tensile strain on MnTe favors the in-plane ferromagnetic coupling and out-of-plane antiferromagnetic coupling as we demonstrate in Appendix B.
The heterostructure of these systems breaks the cubic crystal symmetry.
The FM phases break time-reversal symmetry $T$, but retain either $C_{2x}$ or $C_{2z}$ rotational symmetries depending on whether the magnetization direction is in-plane or out-of-plane, respectively.
On the contrary, the AFM phases break these rotational symmetries but preserve the combination $C_{2x}'=C_{2x}{\cdot}T$ or $C_{2z}'=C_{2z}{\cdot}T$ depending on the N\'eel vector direction, i.e., they exhibit magnetic rotational symmetries.
The latter are relevant because they protect the topological phases described in this work.

\medskip
 \begin{table}[t!]
\caption{Relaxed lattice parameters $a$ and $c$ for the superlattice (HgTe)$_{10}$/(MnTe)$_4$, $a$ corresponds to the optimized lattice parameter of the cubic zinc-blend divided by $\sqrt{2}$ and $c$ is the size of the supercell along the $z$-axis. Difference in total energy for the supercell ($\Delta$E), energy gap (E$_g$) and symmetries for the different magnetic configurations of the (HgTe)$_{10}$/(MnTe)$_4$ superlattice. The zero of the total energy was fixed to the phase AFM2, which is the ground state. The FM phases do not have a gap since they are semimetallic.}
\begin{tabular}{c| c| c| c| c|c}
\hline \hline
\hspace{0.1cm}&\hspace{0.1cm} a (\AA) \hspace{0.1cm}&\hspace{0.1cm} c (\AA) \hspace{0.1cm}& $\Delta$E (meV) & E$_g$ (meV) &  Symmetries \\
\hline
AFM1  &  4.665  & 45.829	& 0.3 	& 47  & $C_{2x}{\cdot}T$ \\
AFM2  &  4.665 	& 45.829	& 0	    & 51  & $C_{2z}{\cdot}T$   \\
FM1   &  4.672	& 45.931    & 123   & SM & $C_{2x}$ \\
FM2   &  4.672	& 45.931    & 124   & SM	& $C_{2z}$ \\
\hline  \hline
\end{tabular} \label{Table1}
\end{table}

\section{Antiferromagnetic phases}
We have performed first-principles calculations for the (HgTe)$_{10}$/(MnTe)$_4$ 3D SLs. We have analysed the magnetic stability evaluating the total energy of different magnetic phases with different spin orientations as shown in Table \ref{Table1}. Looking at the results of the energy difference for the supercell $\Delta{E}$, we observe that the energy difference between in-plane and out-of-plane configurations is 0.3 meV for the AFM phase and 1 meV for the FM phase. These results are compatible with magnetocrystalline anisotropy values for compounds with heavy elements. The exchange interaction is ferromagnetic in the $xy$ plane while it is antiferromagnetic along the $z$ direction. The AFM2 phase is the magnetic ground state. The total energy of the AFM1 phase is slightly higher than the AFM2 phase and much lower than the energy of the FM1 and FM2 phases.
This indicates that the magnetic anisotropy is weaker than the effective magnetic exchange interaction between the Mn atoms in these 3D SLs.

The AFM phases are insulating while the FM phases are semimetallic due to exchange-induced giant spin-splitting of $p$-like bands. This is usual in magnetic systems with magnetic bands at the Fermi level \cite{Autieri_2016}, however, in this particular case, the bands at the Fermi level are bands belonging to the non-magnetic Hg and Te atoms.
However, these atoms hybridize with the magnetic \emph{d}-bands of the Mn acquiring larger effective hopping in the ferromagnetic phases.
Both AFM phases have compensated magnetic moments, such that the total magnetic moment is exactly zero. Compensated antiferromagnets usually do not show an anomalous Hall effect, except in the case of particular symmetries which are not present in zinc-blende MnTe \cite{Smejkal2022}. Therefore, we have zero anomalous Hall effect for the AFM phases considered in this paper.

MnTe has $d^5$ magnetic atoms that couple antiferromagnetically at half-filling due to the superexchange via virtual hopping involving Te anions.
In an ideal cubic structure, the magnetism in MnTe exhibits frustration since the Mn atoms are on an fcc sublattice with twelve nearest neighbors with triangular connectivity \cite{ivanov2017polar}. In the case of in-plane tensile strain, the distance between the Mn atoms in the same layer becomes larger and the in-plane magnetic coupling gets smaller. The AFM coupling between the planes becomes the dominant magnetic interaction and the frustration is lifted. The AFM phase with alternating ferromagnetic layers becomes the magnetic ground state (See Appendix B).

All AFM phases investigated in this paper exhibit no anomalous Hall effect and have zero Chern number.

\begin{figure}
  \begin{center}
        \includegraphics[width=0.99\linewidth]{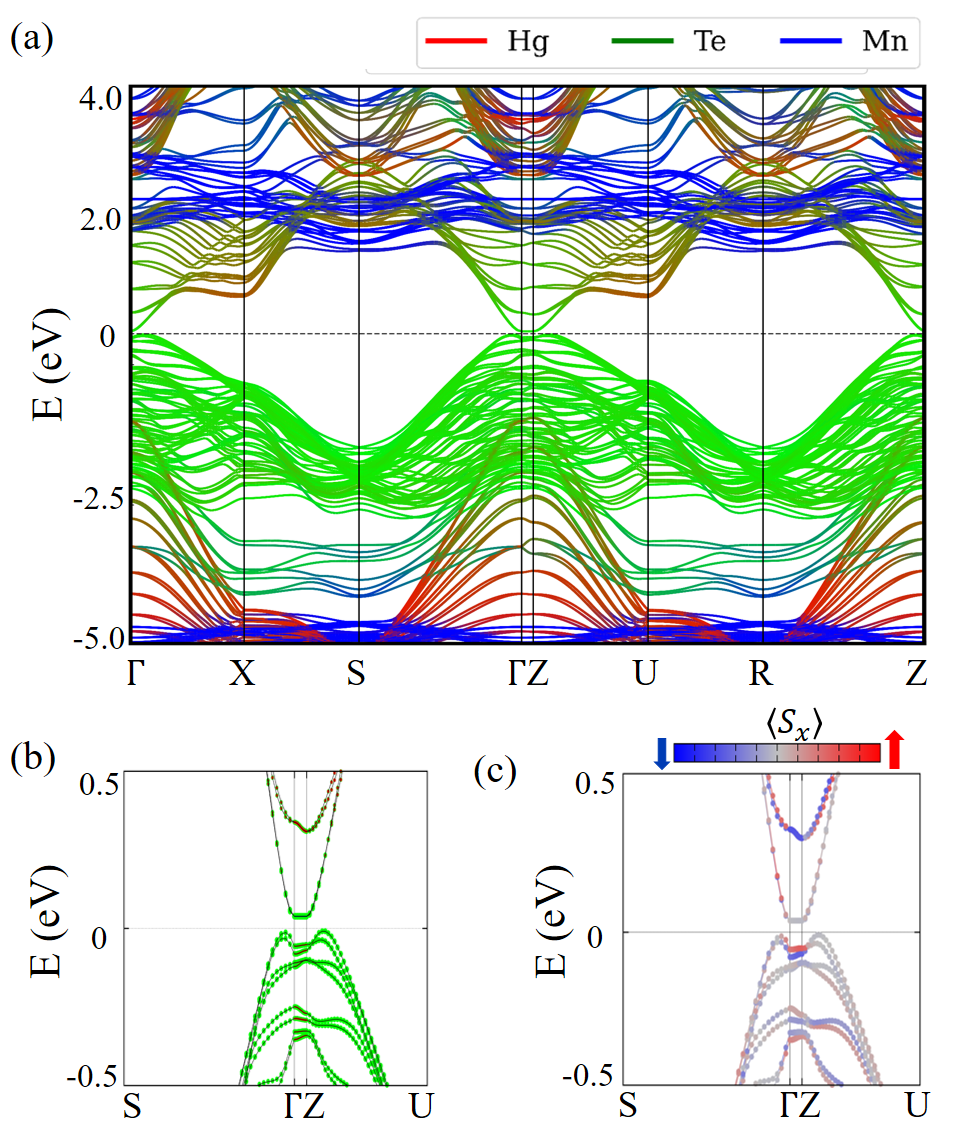}
  \end{center}
   \centering
   \caption{(a) Atomically resolved electronic band structure of the AFM1 configuration in the HgTe/MnTe 3D superlattice in the energy range between -5 eV and +4 eV. (b) Atomically and (c) spin-resolved band structure along the high-symmetry path S-$\Gamma$-Z-U in the energy range between -0.5 eV and +0.5 eV. The Fermi level is set to zero.}
   \label{AFM1_band}
\end{figure}

\begin{figure}
   \includegraphics[width=0.99\linewidth]{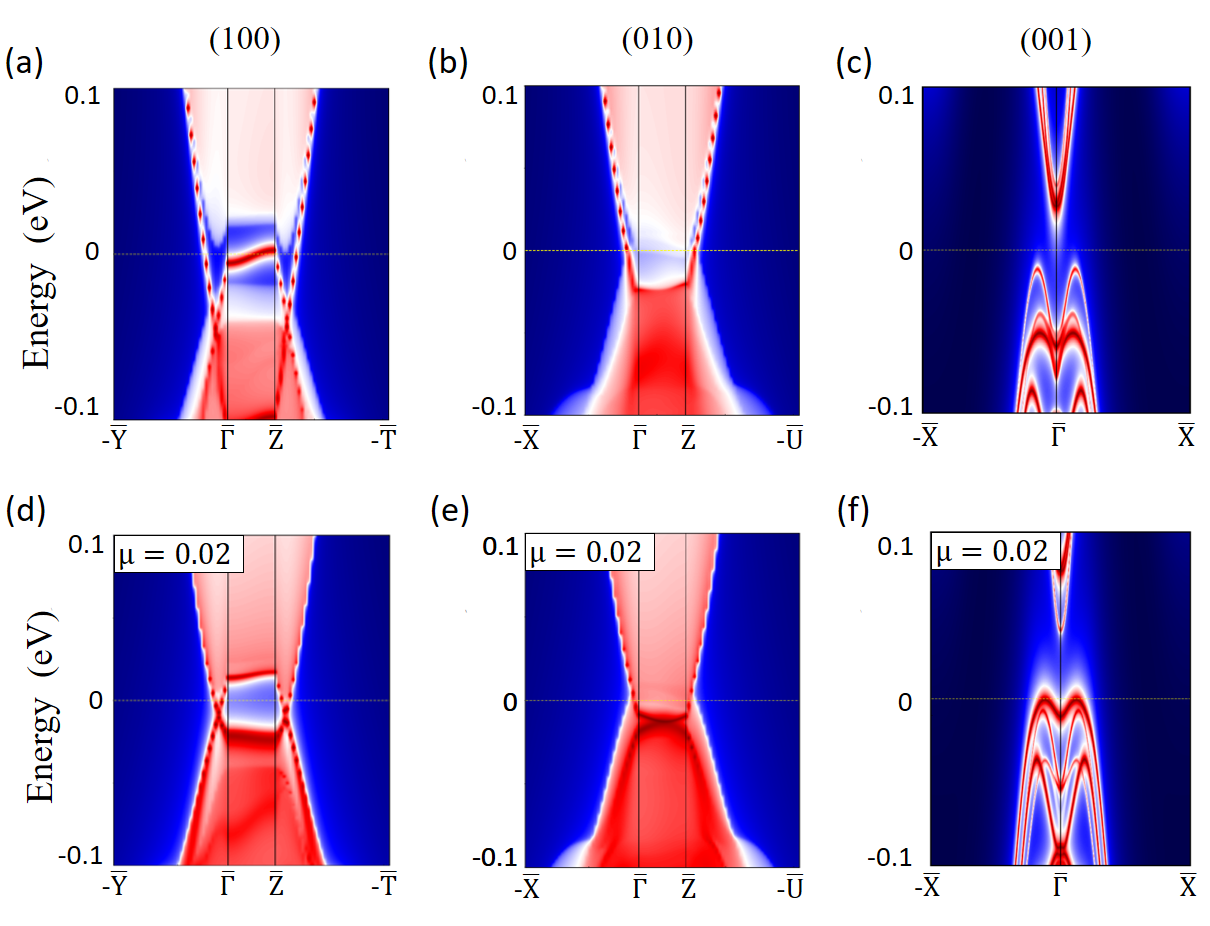}
   \centering
   \caption{Electronic properties of the AFM1 phase with four unit cells of MnTe. Band structure projected on the (a) (100), (b) (010) and (c) (001) surface. Same band structure with an applied on-site potential of 0.02 eV on the surface projected on the (d) (100), (e) (010) and (f) (001) surface. The Fermi level is set to zero. Red means presence of surface electronic states while blue means absence of electronic states, and white is for the bulk states.
   }
   \label{AFM1_surface}
\end{figure}

\subsection{Weak antiferromagnetic topological insulator for 4 MnTe uc with in-plane N\'eel vector}

We first illustrate the AFM topological phase denoted AFM1. The atomically resolved band structure with spin-orbit coupling taken into account is shown in ~\cref{AFM1_band}(a) using the high-symmetry path of the first Brillouin zone [see ~\cref{bulk}(e)]. The bands below the Fermi level are mostly dominated by the \emph{p}-orbitals of the Te-atoms (green bands) except for a single band of Hg \emph{s}-orbitals lying below the Fermi level.
As a consequence, the lowest conduction band has \emph{p}-orbital character signaling the presence of a band inversion at the $\Gamma$ point as shown in ~\cref{AFM1_band}(b).
The bands associated with localized \emph{d}-orbitals of the Mn atoms are far from the Fermi level, which is similar to bulk HgTe doped with Mn \cite{autieri2020momentumresolved}. The top of the valence band along the high-symmetry path $\Gamma$-S shows a camel-back feature which is present in the topological phase of zinc-blende compounds when the bulk crystal symmetry is broken \cite{Cuono22,Islam22,autieri2020momentumresolved}. Another camel-back feature is present in the valence band along Z-U. These camel-back features, along with associated acceptor states \cite{Dietl:2022_arXiv:2206.01613,Dietl:2022_arXiv:2209.03283}, are relevant for the transport properties of HgTe-based systems \cite{yahniuk_app4,Shamin:2020_SA}. The bands are weakly dispersive along the $\Gamma \rightarrow Z$ direction due to the large value of the lattice constant $c$ in the superlattice.
In~\cref{AFM1_band}(c), we observe a Zeeman-like band splitting between spin-up and spin-down states along $x$ at the $\Gamma$ point despite the absence of a net non-zero magnetization.

In~\cref{AFM1_surface}(a), we show the surface band structure projected on the (100) surface with the associated topological surfaces states.
In contrast to topological insulators with time-reversal symmetry, the surface Dirac cones are shifted away from the high-symmetry point for this surface termination but buried inside the bulk bands, as shown in ~\cref{AFM1_surface}(a).
This shift of the Dirac cones is due to the magnetism.
The Dirac cones are pushed up by applying a positive on-site potential $\mu$ on the surface atoms as it was done in~\cref{AFM1_surface}(d-f).
This numerical trick allows us to have a better visualization of the Dirac cones, and to decide whether they are gapped or gapless.
For the (100) case shown in Figs.~\ref{AFM1_surface}(a), we find that the two surface Dirac cones are indeed gapless.
For the (010) surface [see Figs. ~\ref{AFM1_surface}(b,e)], we observe surface-state signatures in agreement with the presence of two Dirac points buried inside the bulk bands, one pinned to $\bar{\Gamma}$ and the other pinned to $\bar{Z}$.
Both Dirac points appear to be gapless.
In contrast to the (100) and the (010) side surfaces, the (001) top/bottom surface does not exhibit any signatures of surface Dirac cones, neither gapless nor gapped as shown in Figs. ~\ref{AFM1_surface}(c,f).

The presence of an even number of gapless Dirac surface states only on the side surfaces of the material indicates that this phase realizes a weak topological phase.
It is a weak phase because the two surface Dirac cones could, in principle, be coupled through the breaking of translational symmetry thereby creating a surface gap.
Therefore, in the AFM1 case, we have a weak magnetic topological insulator with a pair of shifted surface Dirac cones on the (100) surfaces and a pair of pinned surface Dirac cones on the (010) surfaces.
The resulting asymmetry between the surfaces could give rise to unidirectional topological transport properties \cite{PhysRevX.12.021069}.

\begin{figure}[b]
  \begin{center}
        \includegraphics[width=0.99\linewidth]{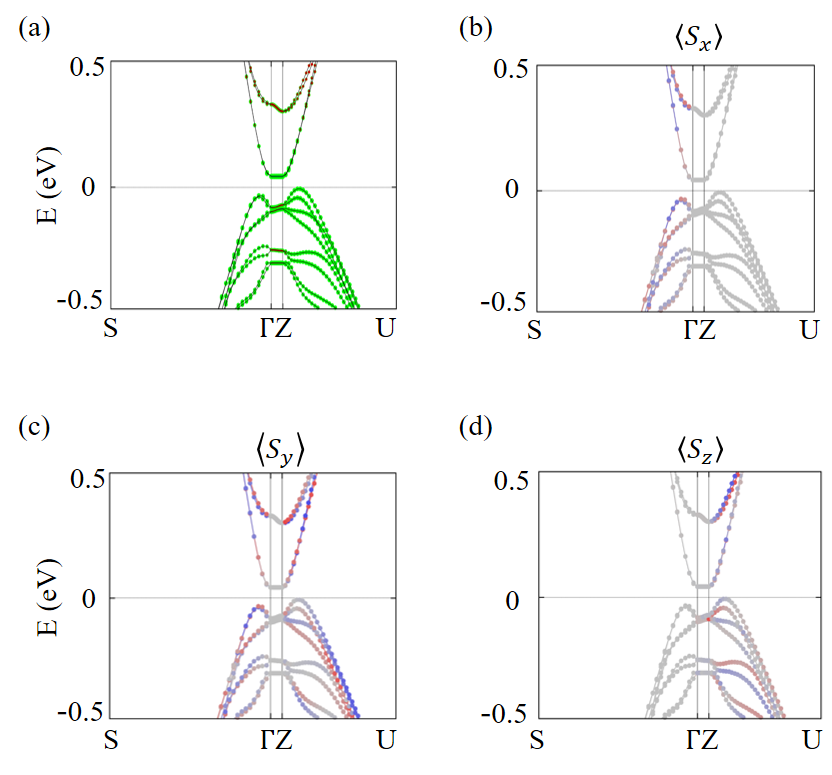}
  \end{center}
   \centering
   \caption{(a) Atomically resolved electronic band structure of the AFM2 phase along the S-$\Gamma$-Z-U high-symmetry path. Tellurium electronic states are represented in green. Band structure resolved in (b) $S_x$, (c) $S_y$ and (d) $S_z$. Legend as in \cref{AFM1_band}(c). The Fermi level is set to zero.}
   \label{AFM2_band}
\end{figure}

\begin{figure}[b]
  \begin{center}
        \includegraphics[width=0.99\linewidth]{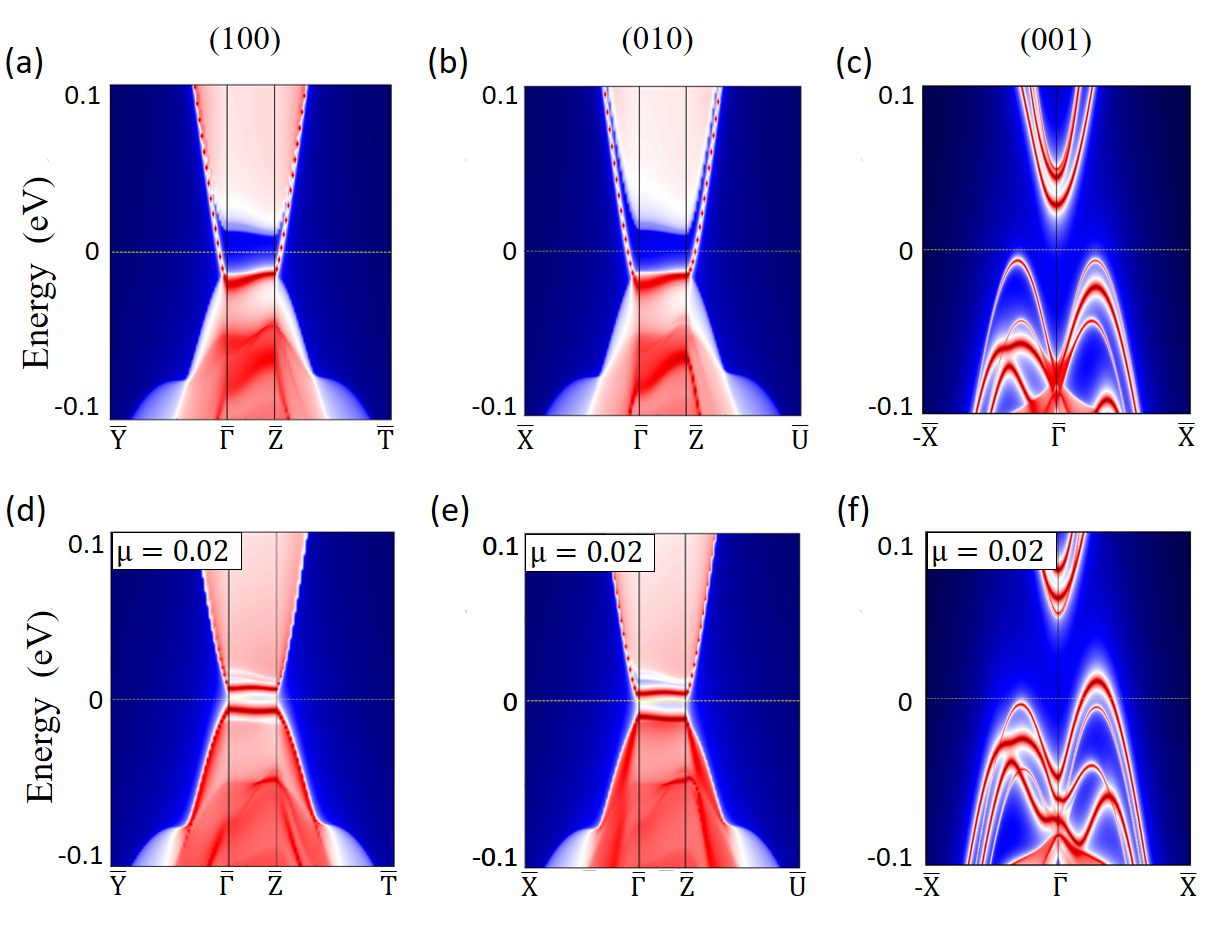}
  \end{center}
   \centering
   \caption{Electronic properties of the AFM2 phase with four unit cells of MnTe. Band structure projected on the (a) (100), (b) (010) and (c) (001) surface. Same band structure with an applied on-site potential of 0.02~eV on the surface projected on the (d) (100), (e) (010) and (f) (001) surface. The Fermi level is set to zero. Red means presence of surface electronic
   states while blue means absence of electronic states, and white is for the bulk states. Once we apply the on-site potential, all surfaces are gapped.}
   \label{AFM2_surface}
\end{figure}

\begin{figure}[b]
  \begin{center}
        \includegraphics[width=0.99\linewidth]{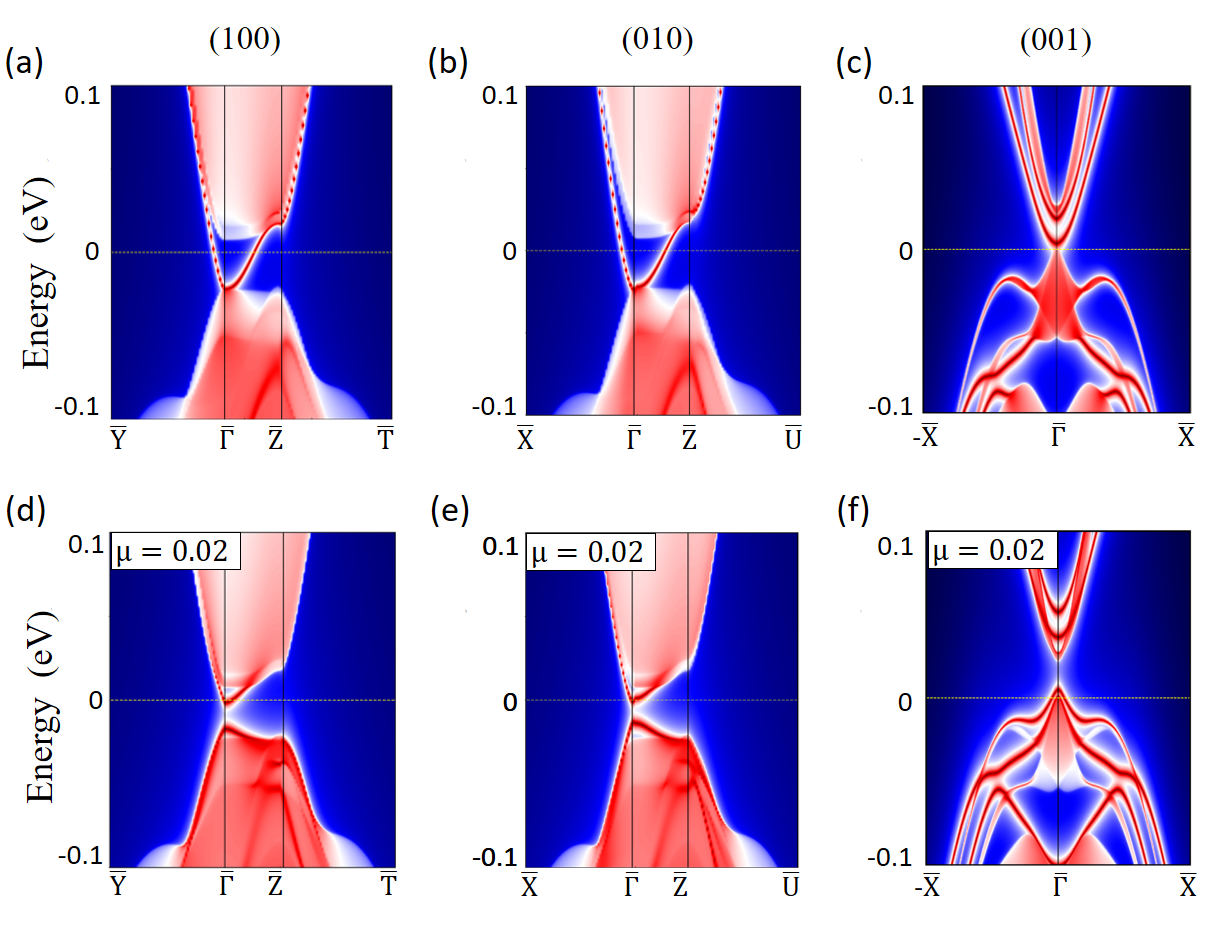}
  \end{center}
   \centering
   \caption{Electronic properties of the AFM2 phase with two unit cells of MnTe. Band structure projected on the (a) (100), (b) (010) and (c) (001) surface. Same band structure with an applied on-site potential of 0.02 eV on the surface projected on the (d) (100), (e) (010) and (f) (001) surface. The Fermi level is set to zero. Red means presence of surface electronic
   states while blue means absence of electronic states, and white is for the bulk states. Once we apply the on-site potential, all surfaces are gapped.}
   \label{AFM2_surface_2uc}
\end{figure}

\subsection{Trivial insulator for 4 MnTe uc with out-of-plane N\'eel vector}

Now, we turn our discussion to the AFM2 phase, corresponding to an out-of-plane N\'eel vector direction, which is the ground state of the HgTe/MnTe 3D SLs. The electronic band structure of the AFM2 phase is quite similar to the AFM1 regarding the bulk bands. The bands near the Fermi level are dominated by the \emph{p}-orbitals of Te while the \emph{d}-orbitals of Mn are far from the Fermi level. Majority-spin bands of Mn are approximately 5 eV below the Fermi level while the minority-spin bands are approximately 2 eV above, as they are for Mn dopants in Te-based zinc-blende structures \cite{autieri2020momentumresolved}. Unlike AFM1, the bands along $\Gamma \rightarrow Z$ are doubly degenerate in AFM2. In ~\cref{AFM2_band}(a-d) we show an enlarged view of the band structure along S$\rightarrow \Gamma \rightarrow $Z $\rightarrow$ U  with orbital and spin projections. Similar to the AFM1 phase, it also has a band inversion at the $\Gamma$ point and camel-back features at the same points.

To inspect the possible topological nature of this phase, we have studied the surface band structure for different atomic terminations.
Figs. \ref{AFM2_surface}(a,b) show the states of (100) and (010) surfaces, respectively.
We find Dirac-like, linearly dispersing surface bands both around $\bar{\Gamma}$ and around the $\bar{Z}$. To see whether the corresponding Dirac cones are gapped or gapless, similar to AFM1, we have applied an on-site potential ($\mu$=0.02 eV) of meV on the surface atoms to move the Dirac cones into the bulk energy gap as we have reported in Figs. \ref{AFM2_surface}(d,e).
We observe a surface energy gap of 9 meV within the bulk energy gap as shown in \cref{AFM2_surface}(d) for the (100) termination.
The (001) surface does not exhibit any signatures of surface Dirac cones, neither gapless or gapped.
Instead, we observe a larger gap reflecting the bulk energy gap as shown in Figs. \ref{AFM2_surface}(c,f).
This is similar to the AFM1 phase.
As there are no gapless surface states on any of the surfaces, the AFM2 phase is a normal insulator.

To understand better the connection between the two phases AFM1 and AFM2, let us assume we could gradually switch off the magnetic moments of the Mn atoms. In this case, the two phases would become identical, because they originally only differed in their N\'eel vector direction. In the process, the surface gaps of the Dirac cones on the (100) and (010) surfaces would close in both phases, while the gap on the (001) surfaces would remain. Also the Dirac cones on the (100) surfaces would move back to the high-symmetry points $\bar{\Gamma}$ and $\bar{Z}$.
The resulting phase would be a weak topological insulator with time-reversal symmetry, which we call the \emph{parent phase} of the magnetic system.

The weak topological nature of the parent phase in the HgTe/MnTe superlattice described above is in contrast to 3D bulk HgTe, which realizes a strong topological insulator under strain. The reason for this difference is the thickness of the MnTe layer, which reduces the coupling between adjacent HgTe layers significantly. Each HgTe layer by itself, therefore, realizes an effective 2D topological insulator. Eventually, the stacking of these 2D TI layers with a weak coupling between them leads to a weak 3D topological phase \cite{Lau2015}.

To make the parent phase a strong topological insulator, we have to reduce the thickness of the Mn layer, which leads to a strong hybridization between the HgTe layers.
A transition from strong TI to weak TI induced by a superlattice with a trivial phase has been observed also for other topological systems \cite{PhysRevB.89.085312}.

\subsection{Axion insulator for 2 MnTe uc with out-of-plane N\'eel vector}

By reducing the thickness of the Mn layer to \emph{two} unit cells, we manage to turn the parent phase of the magnetic system into a strong topological insulator with one surface Dirac cone at $\bar{\Gamma}$ on each surface.

For in-plane magnetization (AFM1), we find a strong magnetic topological phase with one gapless Dirac cone for each of the three surface terminations (see Appendix C). While the Dirac cones are pinned to $\bar{\Gamma}$ for the (010) and (001) surface terminations, the Dirac cone on the (100) surface, which is perpendicular to the in-plane N\'eel vector direction, is unpinned.

For the case of out-of-plane magnetization (AFM2), we find that all surface Dirac cones are gapped out after applying an appropriate surface potential to shift the gapped surface Dirac cones into the bulk energy gap (see \cref{AFM2_surface_2uc}).
Nevertheless, in contrast to the AFM2 phase with four layers of Mn, this phase is nontrivial:
The system preserves the combination of two-fold rotation ($C_{2z}$) and time-reversal symmetry $C_{2z}'=C_{2z}{\cdot}T$, where $C_{2z}$ is the two-fold rotation symmetry which connects the spin-up and spin-down Mn atomic layers.
Due to this symmetry, the quantized nontrivial $\theta$ term ($\theta=\pi$) of the strong topological parent phase remains quantized even in the magnetic phase.
Hence, this phase realizes an axion insulator.
We have checked that the anomalous Hall conductivity near the Fermi level is zero in this phase as expected for an AXI.
The AXI state of HgTe/MnTe is protected by the magnetic $C_{2z}'$ symmetry, therefore it differs from the AXI phase protected by the combination of $T$ and a lattice translation present in other systems \cite{Riberolles2021magnetic}.
To summarize the energetic stability of AXI phase, the AXI is the ground state for 2 uc of MnTe while it disappears for 4 uc of MnTe with the superlattice that becomes trivial. We expect that above 4 uc the AXI will not be stable anymore since the increase in the thickness of the trivial MnTe will push the superlattice deeper into the trivial phase.

The axion insulator phase has been widely investigated in MnTe(Bi$_2$Te$_3$)$_n$ material class \cite{doi:10.1126/sciadv.aaw5685,doi:10.1063/5.0059447,PhysRevLett.122.206401,Liu2020,Klimovskikh2020}.
Therefore, our finding of a critical thickness for the MnTe magnetic layers to obtain the axion insulator in HgTe/MnTe superlattices will also be relevant for MnTe(Bi$_2$Te$_3$)$_n$. We expect that at larger values of $n$ the robustness of the axion insulator will increase although the critical temperature would decrease. Indeed, for $n=3$ a robust axion phase was also achieved~ \cite{PhysRevB.102.045130}.

Reducing the unit cells of the magnetic layers to one is not promising, because the intraplane antiferromagnetic coupling would not lead to a ferromagnetic layer. Therefore, we would not expect an axion insulator in this case.

\section{Ferromagnetic phases}

For ferromagnetically aligned Mn spins, the $p$-$d$ kinetic exchange interaction results in giant splitting of the valence subbands which originate from the Te $p$ states and reside  at the Fermi level. Therefore, the ferromagnetic bulk phases are semimetallic.
The ferromagnetic semimetallic phases of MnTe/HgTe show a large anomalous Hall effect.
Due to the metallicity, the surface states are not always clearly visible.

\subsection{Ferromagnetic semimetal with in-plane magnetic moments}

\begin{figure}
  \begin{center}
        \includegraphics[width=0.99\linewidth]{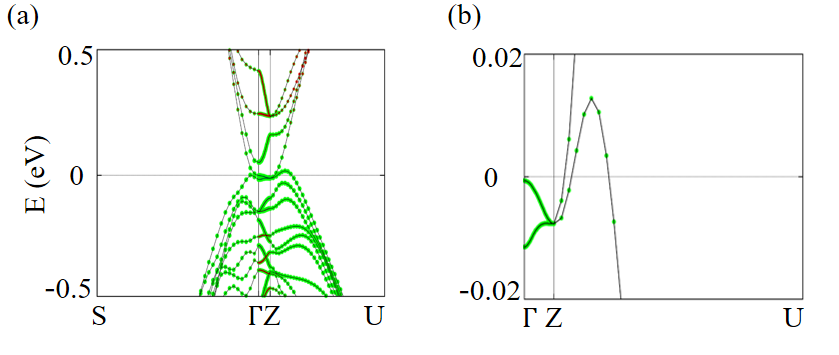}
  \end{center}
   \centering
   \caption{(a) Electronic band structure for the FM1 phase of the HgTe/MnTe 3D SL in the range between -0.5 and +0.5\,eV. (b) Same but in the range between -0.02 and +0.02 eV. The Fermi level is set to zero.}
   \label{FM_INPLANE}
\end{figure}

\begin{figure}
  \begin{center}
        \includegraphics[width=0.99\linewidth]{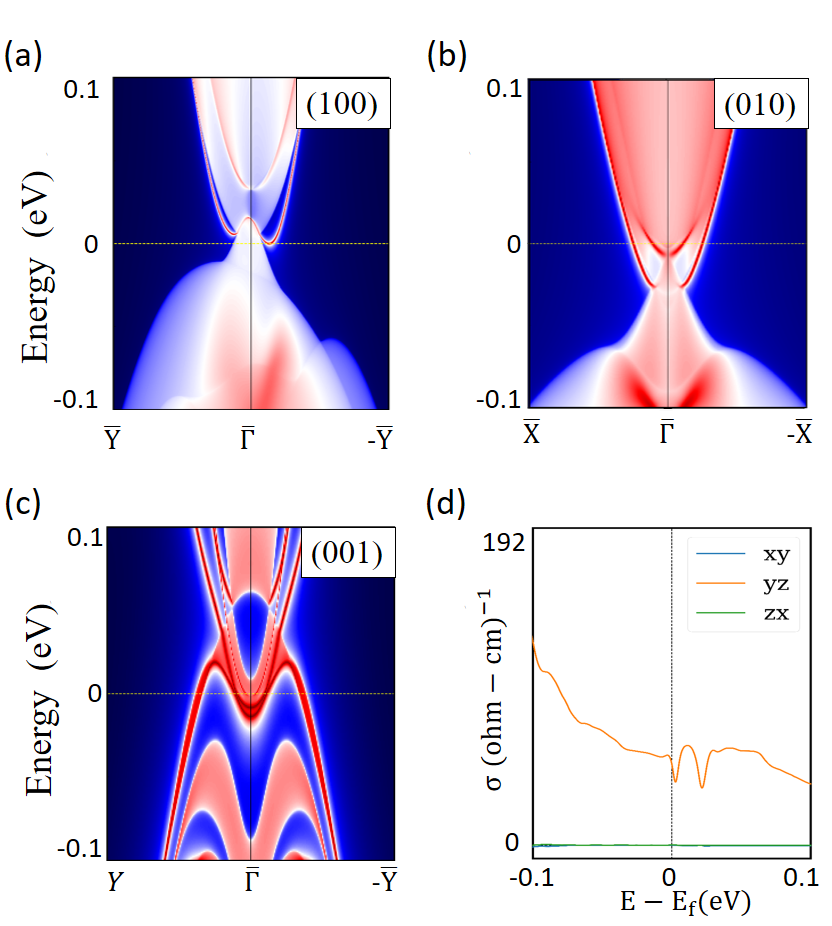}
  \end{center}
   \centering
   \caption{(a-c) Surface band structure projected on the (a) (100), (b) (010) and (c) (001) surface orientation for the FM1 phase of HgTe/MnTe SLs. (d) Anomalous Hall conductivity near Fermi level for the FM1 phase. The Fermi level is set to zero.}
   \label{FM_INPLANE_surface}
\end{figure}

We first discuss the FM1 phase for which spins are aligned along the in-plane $x$ direction [100]. The electronic band structure reveals that the band inversion and the main orbital character of the  bands are the same as in the AFM phases as shown in ~\cref{FM_INPLANE}(a), but the bands are more dispersive along $\Gamma$-Z and the camel-back features are reduced. \cref{FM_INPLANE}(a,b) show the band crossing at the high-symmetry point Z. We observe the coexistence of electron- and hole-pockets typical for a semimetal.
Since the system is close to a topological phase, we also find surface states as shown in ~\cref{FM_INPLANE_surface}(a) (100), ~\cref{FM_INPLANE_surface}(b) (010) and ~\cref{FM_INPLANE_surface}(c) (001) for Te-terminated surfaces.
Furthermore, we observe a nonzero anomalous Hall conductivity $\sigma_{yz}$ shown in ~\cref{FM_INPLANE_surface} (d), which is expected for the $x$ direction of magnetization and a strong spin-orbit interaction for $p$-like subbands.

\subsection{Weyl semimetal with out-of-plane magnetic moments}

\begin{figure*}
  \begin{center}
        \includegraphics[width=0.99\linewidth]{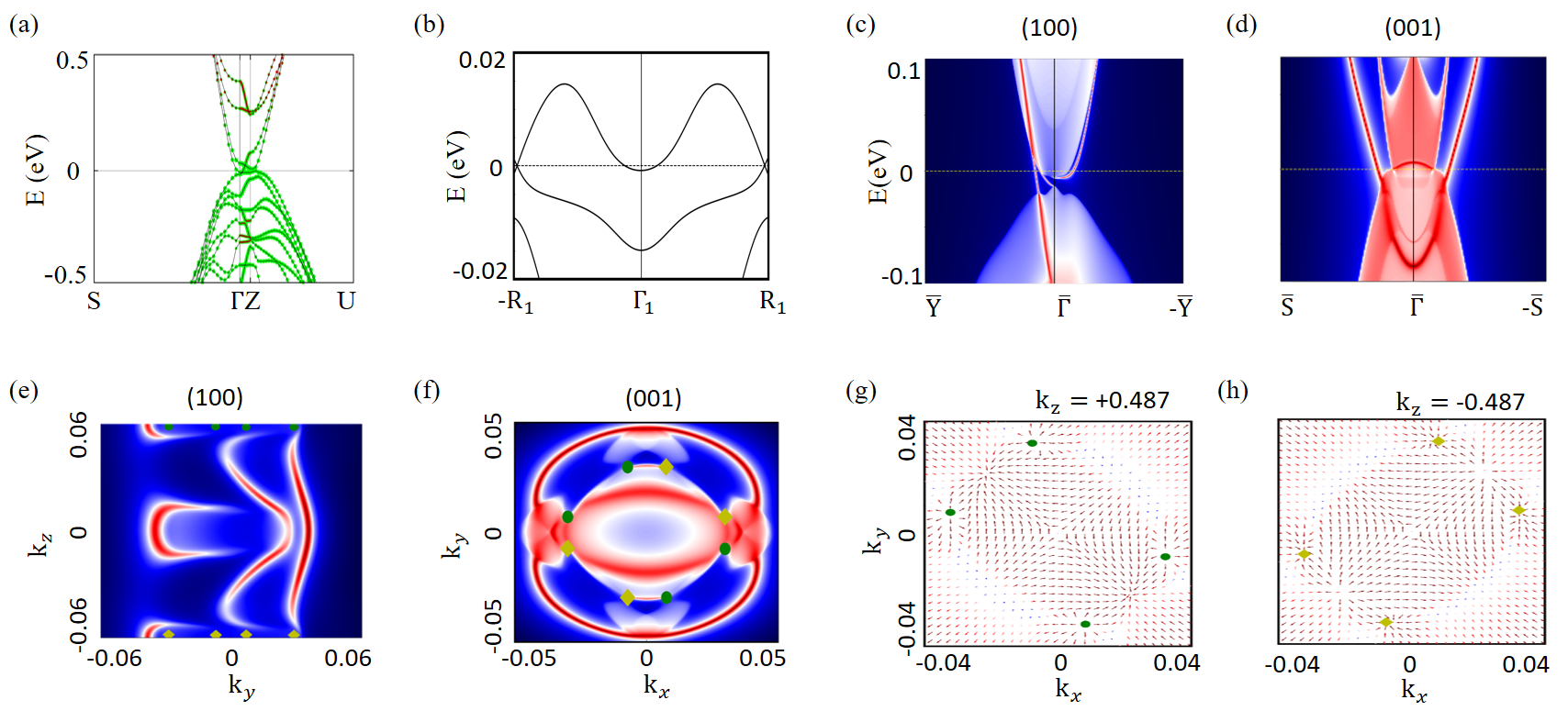}
  \end{center}
   \centering
   \caption{(a) Electronic band structure along S$ \rightarrow \Gamma \rightarrow $Z$ \rightarrow$ U of the HgTe/MnTe 3D superlattice with FM2 magnetic order. (b) Magnified band structure in a specific direction showing the band crossing along the  -R$_1 \rightarrow \Gamma_1 \rightarrow $R$_1$ path with R$_1 =(-0.024\,2\pi/a$,$\pi/a$, $\pi/c$), -R$_1=(-0.024\,2\pi/a$,-$\pi/a$,-$\pi/c$) and $\Gamma_1=(-0.024\,2\pi/a$,0,0). (c,d) Projected surface states on the (100) and (001) surface for the Te-terminated surface. Fermi surface of Te-terminated (e) (100) and (f) (001) surfaces. Green circles and yellow diamonds indicate Weyl points with chirality +1 and -1, respectively. The Weyl points are at the border of the first Brillouin zone along the $k_z$-axis, therefore, the Weyl points with positive and negative chiralities are extremely close. This limits the visualization of their connectivity through the Fermi arcs. We also show the Berry curvature flux in the planes (g) $k_z= +0.487\, 2\pi/c$ and (h) $-0.487 \,2\pi/c$, respectively. The Fermi level is set to zero.}
   \label{FM_OUTOFPLANE}
\end{figure*}

A magnetic Weyl semimetal phase appears with out-of-plane ferromagnetism labelled as FM2. ~\Cref{FM_OUTOFPLANE}(a) shows the atomically resolved band structure along a high-symmetry path of the Brillouin zone.
The bands near the Fermi level are mostly dominated by Te atoms, while the Mn bands are far away from the Fermi level, which is similar to the AFM phases.
In this magnetic configuration, the system shows a semimetallic behavior with four hole-like pockets present in the band structure. The band crossings are not present along the high-symmetry direction.

We observe two-fold degenerate Weyl points at a fixed $\pm k_{x,0}$ or $\pm k_{y,0}$ parallel to the (111) direction as shown in ~\cref{FM_OUTOFPLANE} (b).
The surface bands projected on the (100) and (001) orientations are shown in \cref{FM_OUTOFPLANE}(c) and \cref{FM_OUTOFPLANE}(d), respectively.
The (010) surface results show similar signatures as the (100)  (not shown).

One of the most important signatures of the Weyl semimetals are Fermi arcs connecting Weyl points with opposite chirality on the surface.
The Fermi surface projected on the (100) surface [\cref{FM_OUTOFPLANE}(e)] and on the (001) surface [\cref{FM_OUTOFPLANE}(f)] shows the Weyl points with chirality=+1 (green circle) and chirality=-1 (yellow diamond).
In between, we observe Fermi arcs connecting them.
The Weyl points with chirality +1 are confined in the $k_z=+k_z^*=+0.487\,2\pi/c= 0.0666$\,\AA$^{-1}$ plane with positions ($\sigma_1(k_1+\sigma_2k_2$),$-\sigma_1(k_1-\sigma_2k_2$), $+k_z^*$), whereas the Weyl points with chirality -1 are confined at the $k_z=-k_z^*=-0.487\,2\pi/c= -0.0666$\,\AA$^{-1}$ plane with positions ($\sigma_1(k_1+\sigma_2k_2$),$-\sigma_1(k_1-\sigma_2k_2$),$-k_z^*$), where $\sigma_1$ and $\sigma_2$ take the values $\pm 1$.
The explicit positions of the eight Weyl points are listed in Table \ref{tab2}.
The chirality of the Weyl points is clearly visible in the Berry curvature flux in the $k_z=\pm 0.487\frac{2\pi}{c}$ planes, as reported in ~\cref{FM_OUTOFPLANE}(g-h).
The Weyl points with chirality +1 act as sources of Berry curvature flux, while the Weyl points with chirality -1 represent sinks.

\begin{figure}
  \begin{center}
        \includegraphics[width=0.99\linewidth]{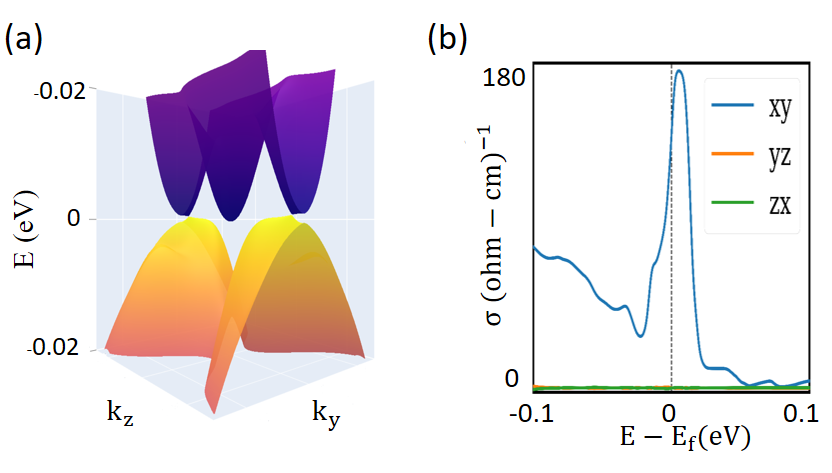}
  \end{center}
   \centering
   \vspace{-0.5cm}
   \caption{(a) Electronic band structure of the FM2 phase at $k_x= 0.006\,2\pi/a$ plane. (b) Anomalous Hall conductivity of the FM2 phase. The Fermi level is set to zero.}
   \label{FM_AQHE}
\end{figure}

The 2D band structure at fixed $k_x$ of the FM2 phase is reported in \cref{FM_AQHE}(a) showing a semimetallic character with two Weyl points.
The finite value of the Chern numbers for momentum-space planes in a range of fixed $k_z$ values gives rise to a large anomalous Hall conductivity $\sigma_{xy}$ in the magnetic Weyl semimetal FM2 as shown in \cref{FM_AQHE}(b).

\begin{table}
  \centering
    \caption{Location of the eight Weyl points (WP) for the phase FM2 and their associated chirality. The four Weyl points with chirality +1 are in the positions ($\sigma_1(k_1+\sigma_2k_2$),$-\sigma_1(k_1-\sigma_2k_2$),$+k_z^*$), whereas the four Weyl points with chirality -1 are in position ($\sigma_1(k_1+\sigma_2k_2)$,$\sigma_1(k_1-\sigma_2k_2$),$-k_z^*$) with $k_1=0.020$\,\AA$^{-1}$, $k_2=0.012$\,\AA$^{-1}$, $\sigma_1=\pm$ and $\sigma_2=\pm$. The units are in \AA$^{-1}$.}\label{tab2}
\begin{tabular}{|c|c|c|c|c|}
\hline
 & \hspace{0.6cm}$k_x$\hspace{0.6cm} & \hspace{0.6cm}$k_y$\hspace{0.6cm} & \hspace{0.6cm}$k_z$\hspace{0.6cm} & Chirality \\ \hline
WP1 & { }0.032   &  -0.008  & { }0.0666   &  { }1  \\
WP2 & -0.032    & { }0.008  &  { }0.0666  &  { }1  \\
WP3 & { }0.008   & -0.032  & { }0.0666   &  { }1  \\
WP4 & -0.008    & { }0.032 & { }0.0666  &  { }1  \\
WP5 & -0.032    & -0.008  &   -0.0666   &    -1  \\
WP6 & { }0.008   & { }0.032  & -0.0666   &    -1  \\
WP7 & { }0.032   & { }0.008  & -0.0666   &    -1  \\
WP8 & -0.008    &  -0.032  & -0.0666   &    -1  \\
\hline
  \end{tabular}
  \newline\newline
\end{table}

Weyl phases can be generated by inversion-symmetry or time-reversal symmetry breaking. In strained HgTe and HgTe/HgSe 3D SL, the Weyl phases were generated by inversion-symmetry breaking in combination with the breaking of the crystal symmetry.
In the presence of trivial layers, such as MnTe or CdTe, we have no Weyl phases generated by inversion-symmetry breaking.

However, the magnetic configuration with out-of-plane magnetic moment gives rise to a ferromagnetic Weyl phase due to time-reversal symmetry breaking. Therefore, the ferromagnetic Weyl phase that we have found is qualitatively different from the Weyl phases in strained HgTe-based heterostructure without time-reversal symmetry breaking. This difference is also reflected in the position of the Weyl points in the two cases.
The positions of the Weyl points generated by the breaking of the crystal symmetry are $(0,\pm k_\parallel^*, \pm k_z^*)$ for the four Weyl points with chirality $+1$, and $(\pm k_\parallel^*, 0, \pm k_z^*)$ for the other four Weyl points with chirality $-1$ as reported in the literature \cite{Islam22,ruan2016ideal}.
Comparing this to the positions of the magnetic Weyl phase (see again Table~\ref{tab2}), we find that the configuration of Weyl points generated by magnetism has a lower symmetry as observed in other material classes \cite{PhysRevB.97.041104}.

\section{Conclusion}

We have engineered an axion insulator phase in MnTe/HgTe zinc-blende 3D superlattices protected by a magnetic rotational symmetry $C_2{\cdot}T$.
The axion insulator phase is observed for the antiferromagnetic order with out-of-plane N\'eel vector direction which is the ground state of the system.
The axion insulator appears for two unit cells of MnTe, a thickness increase of the trivial MnTe weakens the topology producing a trivial insulator for four unit cells of MnTe and above.

For other magnetic configurations, the topology of the superlattice drastically changes to become an antiferromagnetic topological insulator with shifted surface Dirac cones when the direction of the magnetic moments lies in-plane.
The AFM phases have a narrow gap.
Metallicity is introduced into the system for ferromagnetic order due to the giant Zeeman-like splitting of bands in the presence of macroscopic magnetization.
When the magnetization direction switches to in-plane, the ferromagnetic phase transforms into a ferromagnetic semimetal still exhibiting surface states due to the vicinity of a topological phase.
Both ferromagnetic phases show a large anomalous Hall effect close to the Fermi level.

Strong camel-back features appear at the top of the valence band for both antiferromagnetic insulating phases, while they get suppressed in the ferromagnetic semimetallic phases.
We have illustrated how the different magnetic topological phases evolve as a result of changes in the direction of the magnetic moments and in the magnetic order, with the axion insulator phase appearing as the ground state in the specific case of HgTe/MnTe 3D superlattices.
We have demonstrated that it is possible to have an antiferromagnetic axion insulator phase in HgTe-based heterostructures assuming compensated antiferromagnetism with out-of-plane magnetization and two unit cells of MnTe ferromagnetically coupled in the plane.
These conditions are similar to what happens in the MnTe(Bi$_2$Te$_3$)$_n$ material class where the ferromagnetic layers of MnTe are diluted in the Bi$_2$Te$_3$ compound.
The inclusions of the magnetic material should be limited to a few unit cells in order to obtain the axion insulator.
This poses experimental limitations to the creation of new axion insulators, making it difficult to engineer new materials beyond the MnTe(Bi$_2$Te$_3$)$_n$ material class.
More in general, we have provided a recipe to engineer superlattices to obtain new axion insulator phases.

\section*{Acknowledgments}

We thank M. S. Bahramy, W.~Brzezicki, T.~Wojtowicz, V.~V.~Volobuiev and A.~Kazakov for useful discussions. The work is supported by the Foundation for Polish Science through the International Research Agendas program co-financed by the
European Union within the Smart Growth Operational Programme (Grant No. MAB/2017/1).
This work was financially supported by the National Science Center in the framework of the "PRELUDIUM" (Decision No.: DEC-2020/37/N/ST3/02338).
A. L. acknowledges support from a Marie Sk{\l}odowska-Curie Individual Fellowship under grant MagTopCSL (ID 101029345).
C.M.C. was supported by the Swedish Research
Council (VR) through Grant No. 621-2014-4785, Grant No.  2017-04404 and Grant No. 2021-04622.
The work at TIFR Mumbai is supported by the Department of Atomic Energy of the Government of India under project number 12-R$\&$D-TFR-5.10-0100.
We acknowledge the access to the computing facilities of the Interdisciplinary
Center of Modeling at the University of Warsaw, Grant G84-0, GB84-1 and GB84-7.
We acknowledge the CINECA award under the ISCRA initiative IsC93 "RATIO" and IsC99 "SILENTS" grants, for the availability of high-performance computing
resources and support.

\begin{appendices}

\section{Computational details}
The electronic structure calculations were performed within the framework of DFT based on the plane wave basis set and projector augmented wave method using VASP \cite{VASP} package. All calculations are fully relativistic by considering spin-orbit coupling (SOC), a plane-wave energy cut-off of 400~eV has been used. As exchange correlation functional, we use the generalized gradient approximation GGA method \cite{perdew1996generalized} with intra-site Hubbard U for Mn \emph{d}-electrons. We have used the Hubbard U=3.9 eV \cite{PhysRevB.83.184417} and Hund $J_H=0.15U$\, on the \emph{d}-orbitals of the Mn-atoms to estimate more accurate band gap and band order.
The value of $U$ on Mn is usually between 4 and 7\, eV \cite{Keshavarz:2017_PRB,Paul:2014_JAC,Autieri:2014_NJP}. The value that we have used is in the low part of the range, therefore, we are sure that larger values will also provide insulating phases for the ground state. We have  relaxed the geometry of the superlattices.
The crystal structures are fully optimized by relaxing both lattice vectors and atomic positions with force on each atom less than 0.01\,eV/{\AA}. The total energy is converged to 10$^{-8}$\,eV with a
Gaussian-smearing method. We have performed the calculations using 6$\times$6$\times$4 \textbf{k}-points grid centered at the $\Gamma$ point with 144 $k$-points in the Brillouin zone.

We extracted the real space tight-binding Hamiltonian with atom-centred Wannier functions with \emph{s}-Hg, \emph{p}-Te and \emph{d}-Mn orbital projection using the Wannier90 code \cite{mostofi2008wannier90}. We constructed our model using real space tight-binding Hamiltonian to study topological properties using Wanniertools package  \cite{wu2018wanniertools}. The surface energy spectrum was obtained within the iterative Green's function method implemented in Wanniertools \cite{Greenwanniertools}. We evaluate the surface states along all different directions (100), (010) and (001) for the various phases of the system.

\begin{figure}
  \begin{center}
        \includegraphics[width=0.99\linewidth]{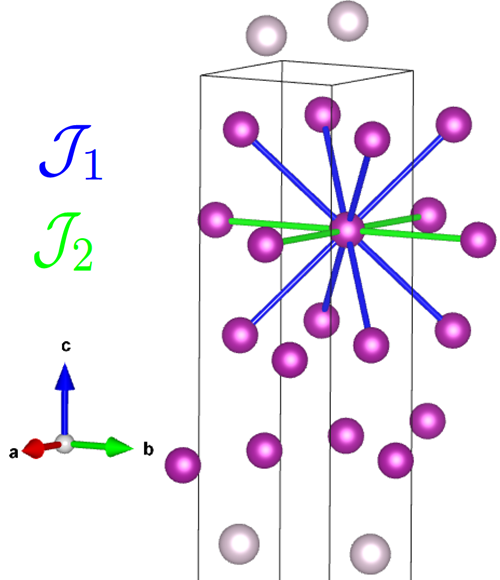}
  \end{center}
   \centering
   \caption{Crystal structure of the MnTe/HgTe superlattice with Mn-Mn bonds highlighted. The grey and purple balls represent the Hg and Mn atoms, respectively. The Mn atoms are shown also outside the supercell. The Te atoms are not shown for a better visualization. The blue and green bonds represent some of the interplane ${\cal{J}}_1$ and intraplane ${\cal{J}}_2$ antiferromagnetic couplings between the Mn atoms. The solid black lines represent the unit cell of the superlattice.}
   \label{Frustration}
\end{figure}

\section{Magnetic properties of the tetragonal M\lowercase{n}T\lowercase{e}}

The sublattice of the Mn atoms in cubic zinc-blende MnTe has an fcc crystal structure. This crystal structure is magnetically frustrated due to the presence of antiferromagnetic couplings in the equilateral triangles composed by two Mn-Mn bonds between atoms in different planes and one Mn-Mn bond between atoms in the same plane.
As shown in Fig.~\ref{Frustration}, we can define ${\cal{J}}_1$ and ${\cal{J}}_2$ as magnetic couplings between Mn atoms of different planes and of the same plane, respectively. In the cubic case, ${\cal{J}}_1$ and ${\cal{J}}_2$ are equal and the system is frustrated.

In the bulk under applied strain or in the superlattice, the cubic system becomes tetragonal and the equilateral triangles become isosceles lifting the degeneracy between ${\cal{J}}_1$ and ${\cal{J}}_2$.
We successfully tested this in the strained bulk MnTe where the energy difference ${\cal{J}}_1-{\cal{J}}_2$ is directly proportional to the strain.
In the case of an anisotropic 2D triangular lattice, we need to have ${\cal{J}}_1/{\cal{J}}_2>2$ to remove the degeneracy and obtain the collinear antiferromagnetic ground state \cite{Schmidt2015-tv}. A similar critical value of the ${\cal{J}}_1/{\cal{J}}_2$ ratio is expected in the fcc lattice.
To have ${\cal{J}}_1/{\cal{J}}_2$ greater than the critical value and stabilize the collinear order, we need a strain of the order of a few percent and an increase of the apical angle of the triangles as previously calculated in other frustrated systems within DFT\cite{Cuono2021Intrachain,Cuono2021Tuning}.

In the MnTe/HgTe superlattice, the angle of isosceles triangles composed by Mn atoms is 61.8$^\circ$ after structural relaxation. Additional strain can increase this ratio and push the ${\cal{J}}_1/{\cal{J}}_2$ ratio above the critical value.
From the experimental point of view, superlattices of MnTe also favors commensurate magnetic order due to the breaking of the crystal symmetry\cite{PhysRevB.48.12817}.
We can safely assume that we can obtain a collinear magnetic ground state with ferromagnetic layers coupled antiferromagnetically in MnTe/HgTe. Therefore, in the main text, we can focus on the topological properties resulting from the collinear phases.

\section{Strong antiferromagnetic topological insulator for 2 M\lowercase{n}T\lowercase{e} uc
with in-plane N\'eel vector}

\begin{figure}
   \includegraphics[width=0.99\linewidth]{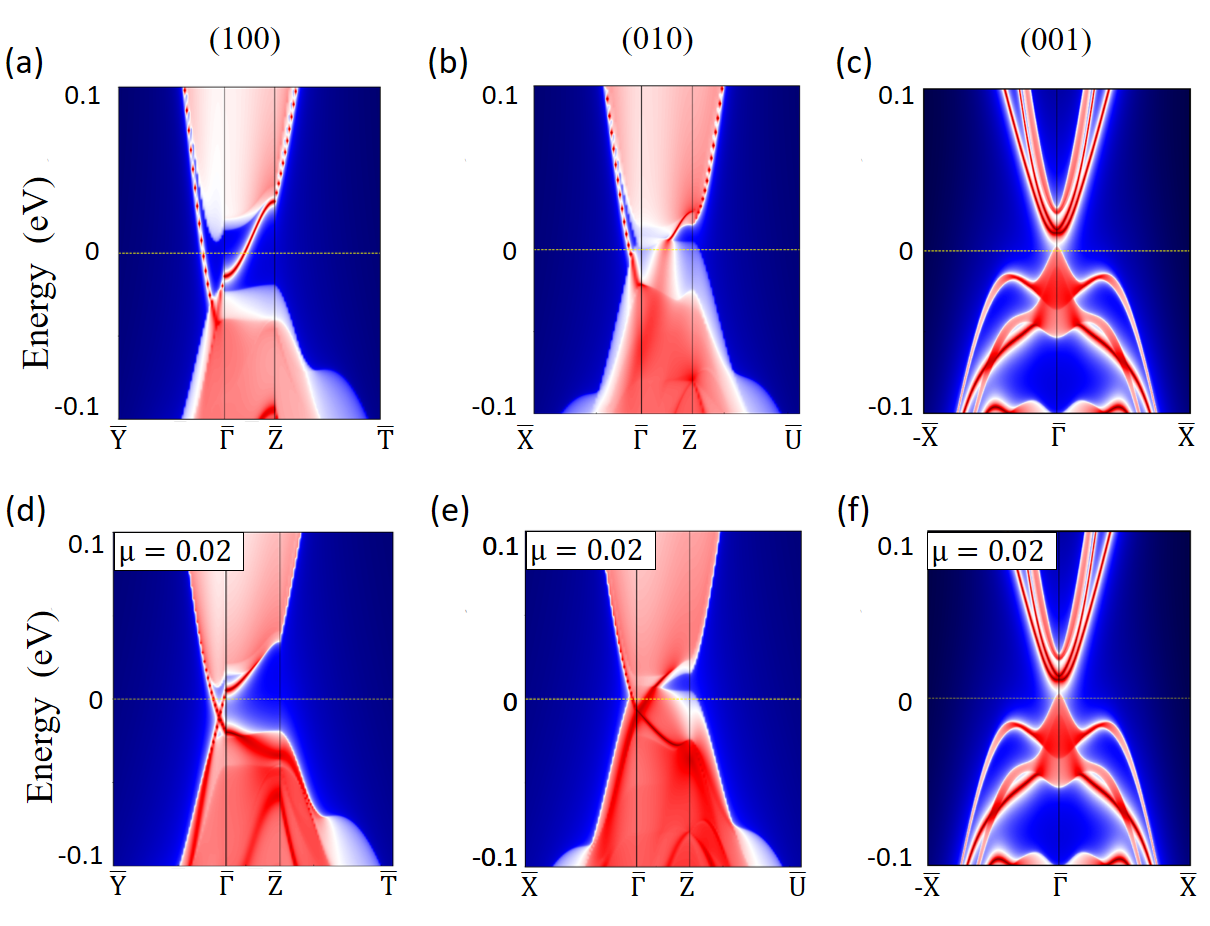}
   \centering
   \caption{Electronic properties of the AFM1 phase with two unit cells of MnTe.
   Band structure projected onto the (a) (100), (b) (010), and (001) surfaces. Band structure of the AFM1 phase with an applied on-site potential of 0.02 eV on the surface projected on the (d) (100) surface, (e) (010) surface, and (f) (001) surface. The Fermi level is set to zero. Red means presence of surface electronic states while blue means absence of electronic states, and white is for the bulk states.}
   \label{AFM1_surface_2uc}
\end{figure}

In \cref{AFM1_surface_2uc}, we report the band structure of the AFM1 phase projected onto the (100), (010), and (001) surfaces without and with an applied on-site potential on the surface atoms.
All three surface terminations exhibit a single gapless surface Dirac cone.
For the (010) and (001) surfaces, the Dirac cones are pinned to the $\bar{\Gamma}$ point, whereas the Dirac cone is shifted along $\bar{\Gamma}$-$\bar{\text{Y}}$ for the (100) termination, which is perpendicular to the in-plane N\'eel vector direction of the magnetic moments.
Due to the observed structure of surface Dirac cones, we infer that this is a strong magnetic topological phase.

\medskip
\newpage

\end{appendices}

\medskip

\bibliography{HgTe_Carmine}

\end{document}